\documentclass[twocolumn,cshowpacs,prb,amsfonts,amsmath,amssymb,floatfix]{revtex4}  

\usepackage{color}
\usepackage{hhline}
\usepackage{mathrsfs}
\usepackage{graphicx}
\usepackage{dcolumn}
\usepackage{bm}
\usepackage{multirow}
\usepackage{booktabs}
\usepackage{ulem}

\input{epsf}

\def\e{\epsilon}
\def\c{\chi}

\begin{document}

\title{Development of a two-particle self-consistent method for multi-orbital systems and its application to unconventional superconductors}

\author{Hideyuki Miyahara$^{1}$}
\author{Ryotaro Arita$^{1,2}$}
\author{Hiroaki Ikeda$^{3}$}
\affiliation{$^1$Department of Applied Physics, University of Tokyo, 7-3-1 Hongo, Bunkyo-ku, Tokyo, 113-8656, Japan} 
\affiliation{$^2$JST PRESTO, Kawaguchi, Saitama, 332-0012, Japan} 
\affiliation{$^3$Department of Physics, Kyoto University, Kyoto 606-8502, Japan} 

\date{\today}

\begin{abstract}
We extend the two-particle self-consistent method proposed by Vilk and Tremblay (J. Phys. I France 7, 1309-1368 (1997)) to study superconductivity in multi-orbital systems. Starting with the sum rules for the spin and charge susceptibilities, we derive self-consistent equations to determine the renormalized effective interactions. We apply this method to the two-orbital $d_{x^2-y^2}$-$d_{3z^2-r^2}$ model for La$_2$CuO$_4$ and the five-orbital $d$-model for LaFeAsO. Comparing the results with those of the random phase approximation or the fluctuation exchange approximation in which vertex corrections are ignored, we discuss how the vertex corrections affect the pairing instability of La$_2$CuO$_4$ and the dominant pairing symmetry of LaFeAsO.
\end{abstract} 
\maketitle 

\section{Introduction}
Since the seminal studies by Suhl\cite{Suhl} and Kondo,\cite{Kondo} 
superconductivity in multi-orbital systems has been one of the major 
topics in condensed matter physics. So far, many kinds of 
multi-orbital superconductors such as MgB$_2$ (Ref.~\onlinecite{MgB2}), 
alkali-doped C$_{60}$ (Ref.~\onlinecite{C60}), Na$_{x}$CoO$_2\cdot y$H$_2$O (Ref.~\onlinecite{takada1}), 
Sr$_2$RuO$_4$ (Ref.~\onlinecite{SROrev}), iron-based superconductors,\cite{ironrev} and 
heavy fermion superconductors\cite{heavySC} have been discovered and studied extensively. 
Theoretically, a variety of exotic unconventional pairing mechanisms going beyond the
Migdal-Eliashberg theory\cite{Schrieffer} have been proposed.
For example, it has been considered for the cobaltate superconductor
that the Hund's coupling (which of course does not exist for single-orbital systems) 
induces triplet superconductivity,\cite{Mochizuki} and it has recently 
become an issue of hot debates whether orbital fluctuations mediate 
superconductivity in the iron-based superconductors.\cite{Kontani}

To investigate these fascinating possibilities, accurate calculations of superconductivity 
in correlated multi-orbital models are indispensable. 
Among many available approaches, 
from the weak coupling side, one often starts with the random phase approximation (RPA). 
Since the pioneering work for the single-band Hubbard model by Scalapino {\it et al.},\cite{scalapino1}
RPA has been successfully applied to various multi-orbital systems.
The fluctuation exchange approximation (FLEX) developed by Bickers {\it et al.},\cite{FLEX}
which includes the self-energy correction self-consistently, has been also widely used. 
Here, the self-energy is calculated in the manner of 
Baym and Kadanoff,\cite{Baym} and conservation laws
for one-particle quantities such as the total energy and momentum are satisfied.
However, due to the absence of vertex corrections, FLEX violates 
conservation laws for two-particle quantities.

Recently, several diagrammatic methods, which take {into} account {some} 
vertex corrections, have been proposed.\cite{kusunose1,onari1}
Among them, the two-particle self-consistent method (TPSC) proposed by Vilk and 
Tremblay\cite{vilk1} is a promising approach in that it is compatible with 
conservation laws in the two-particle level. 
In this method, vertex corrections in the charge and spin channel are assumed to be momentum and frequency independent, 
and they are determined in such a way that the correlation functions meet their sum rules.
With this numerically inexpensive treatment, it has been demonstrated for the single-band Hubbard model 
that TPSC shows good agreement with quantum Monte Carlo (QMC) calculations. 

In this paper, we formulate TPSC for the multi-orbital Hubbard model. 
{First, we} derive a series of equations to determine the vertex correction{s} in the spin and charge 
channel, and then {apply this method} to a two-orbital model
for La$_{2-x}$(Sr/Ba)$_x$CuO$_4$ and 
a five-orbital model for F-doped LaFeAsO. 
Recently, the two-orbital model (which we call the $d_{x^2-y^2}$-$d_{3z^2-r^2}$ model) was studied by FLEX\cite{sakakibara1} to  give an insight into 
the material dependence of superconducting transition temperature ($T_c$). While FLEX successfully 
describes the difference  between La$_{2-x}$(Sr/Ba)$_x$CuO$_4$ ($T_c \sim$ 40 K) and HgBa$_2$CuO$_{4+\delta}$ 
($T_c \sim$ 90 K), it underestimates the pairing instability for La$_{2-x}$(Sr/Ba)$_x$CuO$_4$ 
and $T_c$ is much lower than the experimental value. 
We show that, in the present multi-orbital TPSC calculation, the inter-orbital scattering enhances the $d$-wave instability and
{reasonable value of} $T_c$ is obtained {for the intermediate coupling regime}.
For the five-orbital model, it has been extensively studied by RPA\cite{kuroki1,ironRPA,graser01} and
FLEX.\cite{ironFLEX} There, strong spin fluctuation has been shown to mediate the
$s$-wave superconductivity with sign changes (the {so-called} $s_{\pm}$-wave pairing). 
On the other hand, recently, it has been pointed out that vertex correction{s can} enhance
orbital fluctuations, which mediate $s$-wave superconductivity without
sign changes (the $s_{++}$-wave pairing).\cite{onari1}
{In this paper, we} show that orbital fluctuations are enhanced in TPSC, 
while the dominant pairing symmetry is still $s_{\pm}$ when the system resides in the weak coupling regime.

This paper is organized as follows. In Sec. \ref{sec:method}, we formulate multi-orbital 
TPSC for the Hubbard model. We discuss how we calculate the charge (orbital) and spin 
correlation functions. 
In Sec. \ref{sec:result}, we show the results for the two- and five-orbital Hubbard 
model and the summary of the present study is given in Sec. \ref{sec:summary}.

\section{Method} \label{sec:method}
\subsection{Model} 
The Hamiltonian of the multi-orbital Hubbard model is given by
\begin{align*}
& H=\sum_{{\mathbf r}\mu \sigma}{\epsilon}_{\mu}n_{\mu \sigma}({\mathbf r})+\sum_{{\mathbf r}{\mathbf r'} \mu \nu \sigma} t_{{\mathbf r}{\mathbf r'}}^{\mu \nu} c_{\mu \sigma}^{\dagger}({\mathbf r})c_{\nu \sigma}({\mathbf r'})\\
&+\sum_{\mathbf r} \Bigl[ U\sum_{\mu}n_{\mu \uparrow}({\mathbf r})n_{\mu \downarrow}({\mathbf r}) +U'\sum_{\mu > \nu}\sum_{\sigma {\sigma}'}n_{\mu \sigma}({\mathbf r}) n_{\nu{\sigma}'}({\mathbf r})\\
&-J\sum_{\mu \ne \nu}{\mathbf S}_{\mu}({\mathbf r})\cdot {\mathbf S}_{\nu}({\mathbf r})+J'\sum_{\mu \ne \nu}c_{\mu \uparrow}^{\dagger}({\mathbf r}) c_{\mu \downarrow}^{\dagger}({\mathbf r})c_{\nu \downarrow}({\mathbf r})c_{\nu \uparrow}({\mathbf r}) \Bigr],
\end{align*}
where $c^\dagger_{\mu \sigma}(\mathbf r)$ is a creation operator of an electron with spin $\sigma$ and orbital $\mu$ at site $\mathbf r$, and $n_{\mu \sigma}({\mathbf r})=c_{\mu \sigma}^{\dagger}({\mathbf r})c_{\mu \sigma}({\mathbf r})$, $\mathbf S_{\mu}({\mathbf r})=(c_{\mu \uparrow}^{\dagger}({\mathbf r}), c_{\mu \downarrow}^{\dagger}({\mathbf r}))\,{\boldsymbol \sigma}\,(c_{\mu \uparrow}({\mathbf r}), c_{\mu \downarrow}({\mathbf r}))^T$ with the Pauli matrices $\boldsymbol \sigma$.  
The on-site Coulomb interactions, $U, U', J, J'$ denote the intra-orbital, inter-orbital Coulomb repulsions, the Hund's exchange, and the pair-hopping term, respectively.

\subsection{Two-particle self-consistent method for the single-orbital Hubbard model}
Let us start with a review of TPSC for the single-orbital Hubbard model formulated by Vilk and Tremblay.\cite{vilk1}  The central quantities in this method are the spin and charge correlation functions.  In  the nonmagnetic state, the system holds SU(2) symmetry, and the spin-spin correlation functions do not depend on the spin directions. Thus we consider the $z$ component of the spin operator $S^z({\mathbf r})=n_\uparrow({\mathbf r})-n_\downarrow({\mathbf r})$ and the charge operator $n({\mathbf r})=n_\uparrow({\mathbf r})+n_\downarrow({\mathbf r})$.  In RPA, the spin and charge correlation functions are evaluated as follows:
\begin{equation} \label{eq:RPA}
{\chi}^{\rm sp}_{\rm RPA}(q)=\frac{2{\chi}^0(q)}{1-U{\chi}^0(q)},~~
{\chi}^{\rm ch}_{\rm RPA}(q)=\frac{2{\chi}^0(q)}{1+U{\chi}^0(q)},
\end{equation}
with the irreducible susceptibility,
\begin{equation*}
{\c}^0(q)=-\frac{T}{N}\sum_k G^0(k)G^0(k+q),
\end{equation*}
where $T$ and $N$ are temperature and number of sites in the system, and $G^0(k)=1/({\it i}{\e}_n+\mu-{\e}(\mathbf k))$ is the bare Green's function with chemical potential $\mu$ and energy dispersion $\epsilon(\mathbf k)$.  Here, we have introduced the abbreviations  $k=({\mathbf k},i{\e}_n)$ and $q=(\mathbf q,i{\nu}_n)$ being the fermionic and bosonic Matsubara frequencies, respectively. 

Note that the RPA violates the Pauli principles, and it does not fulfill the following two sum rules,
\begin{subequations}
\label{eq:sumrule}
\begin{align}
\frac{T}{N}\sum_q {\c}^{\rm sp}(q) &=\langle (n_\uparrow({\mathbf r})-n_\downarrow({\mathbf r}))(n_\uparrow({\mathbf r})-n_\downarrow({\mathbf r})) \rangle \nonumber\\
&=n-2\langle n_{\uparrow}n_{\downarrow}\rangle, \\
\frac{T}{N}\sum_q {\c}^{\rm ch}(q) &=\langle (n_\uparrow({\mathbf r})+n_\downarrow({\mathbf r}))(n_\uparrow({\mathbf r})+n_\downarrow({\mathbf r})) \rangle-n^2 \nonumber\\
&=n+2\langle n_{\uparrow}n_{\downarrow}\rangle -n^2 , 
\end{align}
\end{subequations}
which are exact relations derived via the Pauli principles, $\langle n_{\sigma}({\mathbf r})^2\rangle =\langle n_{\sigma}({\mathbf r})\rangle $ (see Appendix \ref{sec:appendix1}). Here, $n$ is the particle number per site and for nonmagnetic states, $\langle n_{\uparrow}({\mathbf r})\rangle =\langle n_{\downarrow}({\mathbf r})\rangle =n/2$. Note that the double occupancy, $\langle n_\uparrow({\mathbf r})n_\downarrow({\mathbf r}) \rangle \equiv \langle n_\uparrow n_\downarrow \rangle$  is  also translation invariant and does not depend on site $\mathbf r$.

In TPSC, to meet the above conditions [Eqs. (\ref{eq:sumrule})], we introduce two independent effective interactions, $U^{\rm sp}$ for the spin channel and $U^{\rm ch}$ for the charge channel.  Then the full susceptibilities of Eq. (\ref{eq:RPA}) are replaced with
\begin{equation} \label{eq:chi}
{\chi}^{\rm sp}(q)=\frac{2{\chi}^0(q)}{1-U^{\rm sp}{\chi}^0(q)},~~
{\chi}^{\rm ch}(q)=\frac{2{\chi}^0(q)}{1+U^{\rm ch}{\chi}^0(q)}.
\end{equation}

Finally, we put the following ansatz:
\begin{eqnarray}
U^{\rm sp}=\frac{\langle n_{\uparrow}n_{\downarrow}\rangle }{\langle n_{\uparrow}\rangle \langle n_{\downarrow}\rangle }U, \label{eq:ansatz} 
\end{eqnarray}
which is compatible with the equations of motion (see Appendix \ref{sec:appendix2}).
Equations (\ref{eq:sumrule}), (\ref{eq:chi}), and (\ref{eq:ansatz}) provide a set of self-consistent equations in TPSC. Namely, $U^{\rm ch}$, $U^{\rm sp}$, and $\langle n_{\uparrow}n_{\downarrow}\rangle $ are self-consistently determined for given $n$ and $U$.
The one-particle Green's function and self-energy are calculated by 
\begin{align*}
G(k) & =G^0(k)+G^0(k)\Sigma(k)G(k), \\
\Sigma(k) &= \frac{1}{4}\frac{T}{N}\sum_q \Bigl[ U^{\rm sp}\chi^{\rm sp}(q)U
+U^{\rm ch}\chi^{\rm ch}(q)U \Bigr] G(k-q).
\end{align*}
For the single-band Hubbard model, it has been demonstrated that TPSC agrees well with QMC.\cite{vilk1}

\subsection{Extension to multi-orbital systems}
Let us here formulate TPSC for the multi-orbital Hubbard model.
{Hereafter, we follow the matrix form employed in Refs. \onlinecite{Mochizuki} and \onlinecite{takimoto1}.}
The irreducible susceptibility is defined as 
\begin{eqnarray}
{\chi}^0_{\lambda \mu \nu \xi}(q)=-\frac{T}{N}\sum_{k } G^0_{\nu \lambda}(k)G^0_{\mu \xi}(k+q),
\end{eqnarray} 
which can be considered as a matrix element with a row $\lambda\mu$ and a column $\nu\xi$ of a matrix ${\boldsymbol \chi}^0(q)$.
In the nonmagnetic state, the system is invariant for spin rotation, and then $2{\boldsymbol \chi}^{{\rm sp}z}(q)={\boldsymbol \chi}^{{\rm sp}\pm}(q)$ holds, where ${\boldsymbol \chi}^{{\rm sp}\pm}(q)$ is the in-plane correlation function between $S^\pm_{\lambda\mu}({\mathbf r})=S^x_{\lambda\mu}({\mathbf r})\pm iS^y_{\lambda\mu}({\mathbf r})=c_{\lambda\sigma}^{\dagger}({\mathbf r})c_{\mu\bar\sigma}({\mathbf r})$ with $\sigma=\uparrow$ or $\downarrow$.
In TPSC, similar to the single-orbital case, the spin and charge susceptibilities are given by
\begin{subequations}
\label{eq:boldchi}
\begin{align}
{\boldsymbol \chi}^{\rm sp}(q)& =({{\mathbf 1}-{\boldsymbol \chi}^0}(q){\mathbf U}^{\rm sp})^{-1}{2{\boldsymbol \chi}^0(q)}, \\
{\boldsymbol \chi}^{\rm ch}(q)& =({{\mathbf 1}+{\boldsymbol \chi}^0}(q){\mathbf U}^{\rm ch})^{-1}{2{\boldsymbol \chi} ^0(q)}, 
\end{align}
\end{subequations}
where ${\mathbf U}^{\rm sp(ch)}$ is the renormalized effective interaction matrix for  the spin (charge) channel.\cite{takimoto1,Mochizuki}  
For two-orbital systems, for instance, these are represented as
\begin{subequations}
\begin{align}
{\mathbf U}^{\rm sp}&=
\left(
\begin{array}{cccc}
U^{\rm sp}_{1111} & J^{\rm sp} & 0 & 0 \\
J^{\rm sp} & U^{\rm sp}_{2222} & 0 & 0 \\
0 & 0 & {U}^{\rm sp}_{1212} & J^{\rm sp} \\
0 & 0 & J^{\rm sp} & {U}^{\rm sp}_{2121} 
\end{array}
\right), \\
{\mathbf U}^{\rm ch}&=
\left(
\begin{array}{cccc}
U^{\rm ch}_{1111} & \hspace*{-5pt} 2{U}^{\rm ch}_{1122}-J^{\rm ch} & 0~~ & 0~~ \\
2{U}^{\rm ch}_{2211}-J^{\rm ch} & U^{\rm ch}_{2222} & 0~~ & 0~~ \\
0 & 0 & \hspace*{-12pt} -{U}^{\rm ch}_{1212}+2J^{\rm ch} & 0~~ \\
0 & 0 & 0~~ & \hspace*{-12pt} -{U}^{\rm ch}_{2121}+2J^{\rm ch}
\end{array}
\right),
\end{align}
\end{subequations}
where $U^{\rm sp}_{\mu \mu \mu \mu}$ ($U^{\rm ch}_{\mu \mu \mu \mu}$) is the intra-orbital Coulomb interaction; $U^{\rm sp}_{\mu \nu \mu \nu}$ ($U^{\rm ch}_{\mu \nu \mu \nu}$) with $\mu \neq \nu$, the inter-orbital Coulomb interaction; $J^{\rm sp}$ ($J^{\rm ch}$), the Hund's coupling.\cite{takimoto1,Mochizuki} In the present study, for simplicity, we ignore the Hund's coupling in the charge channel, namely, $J^{\rm ch}=0$.\cite{comment}  In RPA, one employs the unperturbed bare vertex as follows, $U^{\rm sp (ch)}_{\mu\mu\mu\mu}=U$, $U^{\rm sp (ch)}_{\mu\mu\nu\nu}=U^{\rm sp (ch)}_{\mu\nu\mu\nu}=U'$, and $J^{\rm sp (ch)}=J$.

Next let us consider the sum rule for multi-orbital systems (see Appendix \ref{sec:appendix1}).  In the $n$-orbital Hubbard model, there are $n^4$ sum rules for ${\boldsymbol \chi}^{\rm sp}(q)$ and ${\boldsymbol \chi}^{\rm ch}(q)$. Among them, we use the following equations to determine ${\mathbf U}^{\rm sp}$: 
\begin{subequations}
\label{eq:rule1}
\begin{eqnarray}
\frac{T}{N}\sum_q {\chi}^{{\rm sp}z}_{\mu \mu\mu \mu}(q)&=&2\langle n_{\mu \uparrow}\rangle -2\langle n_{\mu \uparrow}n_{\mu \downarrow}\rangle, \\
\frac{T}{N}\sum_q {\chi}^{{\rm sp}\pm}_{\mu \nu \mu \nu}(q)&=&2\langle c^{\dagger}_{\mu \uparrow}c_{\nu \downarrow}c^{\dagger}_{\nu \downarrow}c_{\mu \uparrow}\rangle \nonumber \\
&=&2\langle n_{\mu \uparrow}\rangle -2\langle n_{\mu \uparrow}n_{\nu \downarrow}\rangle, \\
\frac{T}{N}\sum_q {\chi}^{{z}\pm}_{\mu \mu\nu \nu}(q)&=&2\langle n_{\mu \uparrow} n_{\nu \uparrow}\rangle -2\langle n_{\mu \uparrow}n_{\mu \downarrow}\rangle. 
\end{eqnarray}
\end{subequations}
Note that the intra-orbital component of the sum rule has the same form as that of the single-orbital Hubbard model. For the inter-orbital components, we use ${\boldsymbol \chi}^{{\rm sp}\pm}(q)$ rather than ${\boldsymbol \chi}^{{\rm sp}z}(q)$, since they can be expressed in terms of the density operators.

For ${\mathbf U}^{\rm ch}$, we use the following sum rules for the charge susceptibilities which can be represented by the spin susceptibilities and the double occupancy:
\begin{subequations}
\label{eq:rule2}
\begin{eqnarray}
\frac{T}{N}\sum_q {\chi}^{\rm ch}_{\mu \mu \mu \mu}(q)&=&\langle (n_{\mu \uparrow}+n_{\mu \downarrow})(n_{\mu \uparrow}+n_{\mu \downarrow})\rangle-\langle n_\mu \rangle \langle n_\mu \rangle, \nonumber \\
&=& n_\mu+2\langle n_{\mu\uparrow}n_{\mu\downarrow} \rangle - n_\mu^2 \\
\frac{T}{N}\sum_q {\chi}^{\rm ch}_{\mu \nu \mu \nu}(q)&=&\langle (c^{\dagger}_{\mu \uparrow}c_{\nu \uparrow}+c^{\dagger}_{\mu \downarrow}c_{\nu \downarrow})(c^{\dagger}_{\nu \uparrow}c_{\mu \uparrow}+c^{\dagger}_{\nu \downarrow}c_{\mu \downarrow})\rangle \nonumber \\
&=&\langle n_{\mu \uparrow}(1-n_{\nu \uparrow})\rangle +\langle n_{\mu \downarrow}(1-n_{\nu \downarrow})\rangle \nonumber \\
&&+\langle c^{\dagger}_{\mu \downarrow}c_{\nu \downarrow}c^{\dagger}_{\nu \uparrow}c_{\mu \uparrow}\rangle 
+\langle c^{\dagger}_{\mu \uparrow}c_{\nu \uparrow}c^{\dagger}_{\nu \downarrow}c_{\mu \downarrow}\rangle \nonumber \\
&=&\frac{T}{N}\biggl(\sum_q {\chi}_{\mu \nu \mu \nu}^{{\rm sp}z}(q)+2{\chi}_{\mu \mu\nu \nu}^{{\rm sp}z}(q)\biggr). 
\end{eqnarray}
\end{subequations}

Finally, as in the single-band case, we introduce the following ansatz between the two-particle quantities and the interaction parameters  (see Appendix \ref{sec:appendix3});
\begin{subequations}
\label{eq:re00}
\begin{eqnarray}
U^{\rm sp}_{\mu \mu \mu \mu}&=&\frac{\langle n_{{\sigma} \mu}n_{\bar{\sigma} \mu}\rangle }{\langle n_{{\sigma} \mu}\rangle \langle n_{\bar{\sigma} \mu}\rangle }U, \ \label{eq:re1} \\
{U}^{\rm sp}_{\mu \nu \mu \nu}&=&\frac{\langle n_{{\sigma} \mu}n_{\bar{\sigma} \nu}\rangle }{\langle n_{{\sigma} \mu}\rangle \langle n_{\bar{\sigma} \nu}\rangle }U',\label{eq:re2} \\
{U}^{\rm sp}_{\mu \nu \mu \nu}-J^{\rm sp}&=&\frac{\langle n_{{\sigma} \mu}n_{{\sigma} \nu}\rangle }{\langle n_{{\sigma} \mu}\rangle \langle n_{{\sigma} \nu}\rangle }(U'-J)\label{eq:re3}. 
\end{eqnarray}
\end{subequations}
Equations.~(\ref{eq:boldchi})-(\ref{eq:re00}) are a set of self-consistent equations in the multi-orbital case. 

\subsection{Eliashberg equation}
Superconductivity has been studied by the following linearized Eliashberg equation,
\begin{equation}
\begin{split}
\lambda {\Delta}_{ll'} (k) &= \sum_{k',m_i} V_{lm_1m_4l'}(k ,k') {G}_{m_1m_2}(k') \\
 &\hspace{50pt} \times{\Delta}_{m_2m_3}(k') {G}_{m_4m_3}(-k').\label{gapeq1}
\end{split}
\end{equation}
Eigenstate $\Delta_{ll'}(k)$ with the largest eigenvalue $\lambda$
was numerically evaluated by the power method.  The superconducting transition occurs at the temperature for which $\lambda$ becomes unity.  Here, $G_{ll'}(k)$ is the dressed Green's function,
\begin{equation}
G_{ll'}(k)=G^0_{ll'}(k)+G^0_{lm}(k)\Sigma_{mm'}(k)G_{m'l'}(k),
\end{equation}
and the self-energy $\Sigma_{ll'}(k)$ is given by
\begin{equation}
\begin{split}
\Sigma_{ll'}(k) &
=\frac{1}{4}\frac{T}{N}\sum_q \Bigl[ {\mathbf U}^{\rm sp}{\boldsymbol \chi}^{\rm sp}(q){\mathbf U}_0^{\rm sp} \\
&+{\mathbf U}^{\rm ch}{\boldsymbol \chi}^{\rm ch}(q){\mathbf U}_0^{\rm ch} \Bigr]_{lml'm'} G_{mm'}(k-q).
\end{split}
\end{equation}
In the present study, we omit the Hartree-Fock term, since a part of its contribution is already considered in the one-body part of the Hamiltonian, which is derived from density functional calculation.

The effective interaction $V_{ll'mm'}(k ,k')$ for the spin-singlet pairing can be expressed in a matrix form as follows:
\begin{equation}
\begin{split}
{\mathbf V}(k)& =-\frac{3}{2} {\mathbf U}^{\rm sp}{\boldsymbol \chi}^{\rm sp}(k)\  {\mathbf U}^{\rm sp}_0 +\frac{1}{2} {\mathbf U}^{\rm ch} {\boldsymbol \chi}^{\rm ch}  (k) {\mathbf U}^{\rm ch}_0 \\
& ~~~~~~~ -\frac{1}{2} {\mathbf U}^{\rm sp}_0-\frac{1}{2} {\mathbf U}^{\rm ch}_0 \label{eliash-eq},
\end{split}
\end{equation}
where ${\mathbf U}^{\rm sp}_0$ and ${\mathbf U}^{\rm ch}_0$ are the bare vertex in the spin and charge channel, respectively.\cite{takimoto1,Mochizuki}

\section{Results} \label{sec:result}
Let us move on to the application of the multi-orbital TPSC method to the effective models for $\rm La_{2}CuO_4$ and $\rm LaFeAsO$. Using the technique of the maximally localized Wannier functions,\cite{marzari1} these models are derived from first-principles calculation{s}. In the density-functional calculation{s}, we employed the exchange correlation functional proposed by Perdew {\it et al.}, \cite{Perdew} and the augmented plane wave and local orbital (APW+lo) method as implemented in the WIEN2K program.\cite{WIEN2k}  We then constructed the Wannier functions for the $d$ bands around the Fermi level, using the WIEN2Wannier (Ref.~\onlinecite{w2w}) and the wannier90 (Ref.~\onlinecite{wannier90}) codes.

\subsection{$\rm La_{2}CuO_{4}$}
Recently, the two-orbital $d_{x^2-y^2}$-$d_{3z^2-r^2}$ Hubbard model for the cuprates were studied to understand the material dependence of $T_c$ by 
FLEX.\cite{sakakibara1} There,  the energy difference between the $d_{x^2-y^2}$ orbital and the $d_{3z^2-r^2}$ orbital was found to be {a} key parameter to characterize
La$_2$CuO$_4$ and HgBa$_2$CuO$_4$. Namely, in the FLEX calculation for the two-orbital model, the pairing instability is stronger {in} the latter. 
On the other hand, {in} the former,  
the eigenvalue of the Eliashberg equation within FLEX does not reach unity down to $T\sim$ 40 K. 
The purpose of this subsection is to examine how the vertex corrections in TPSC affect the superconductivity in La$_2$CuO$_4$.

\begin{figure}[htbp]
  \begin{center}
   \includegraphics[width=65mm]{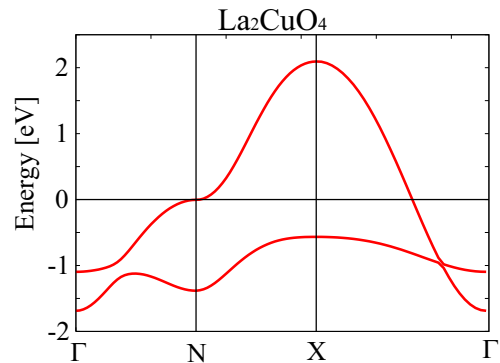}
  \end{center}
 \caption{(Color online) Band structure of the two-orbital model for $\rm La_2CuO_4$. The model consists of the $d_{x^2-y^2}$ orbital and the $d_{3z^2-r^2}$ orbital. 
  The Fermi level is set at $0$ eV. }
   \label{fig1}
\end{figure}

The band structure of the effective two-orbital model for La$_2$CuO$_4$ is shown in Fig. \ref{fig1}. 
We set {$U=2.0$ eV, $U'=1.6$ eV, $J=0.2$ eV}, and $n=2.85$. We employ 64 $\times$ 64 $k$-point meshes and 2048 Matsubara frequencies.
 Hereafter, orbitals 1 and 2 denote the $d_{x^2-y^2}$ and $d_{3z^2-r^2}$ orbitals, respectively.{\cite{note}}

\begin{figure}[htbp]
  \begin{center}
     \includegraphics[width=65mm]{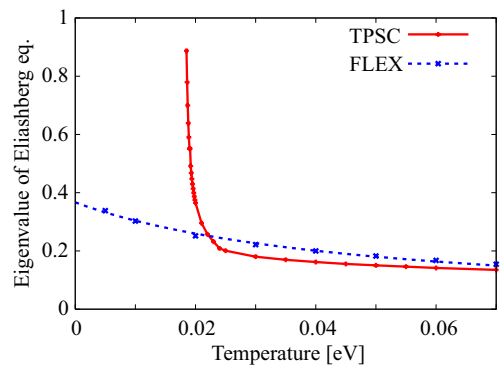}
  \end{center}
\caption{(Color online) Temperature dependence of the maximum eigenvalue of the linearized Eliashberg equation obtained by TPSC (red solid line) and FLEX (blue dotted line). $U$ and $J$ are 2.0 eV and 0.2 eV, respectively, and $n$ is set to be 2.85.}
\label{fig2}
\end{figure}

\begin{figure*}[htbp]
     \includegraphics[width=0.8\textwidth]{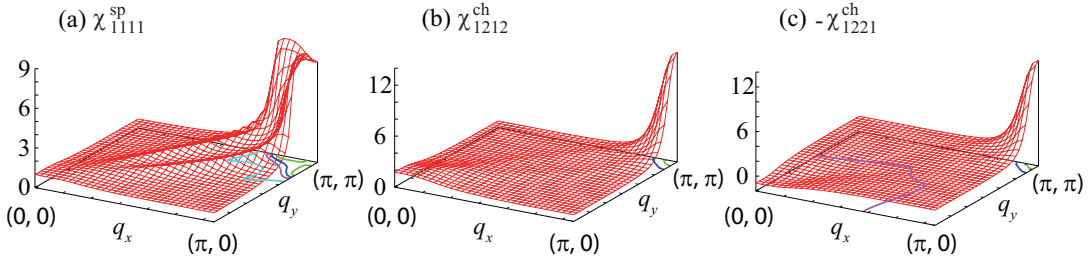}
 \caption{(Color online) (a) ${\chi}^{\rm sp}_{1111}({\mathbf q},\omega =0)$, (b) ${\chi}^{\rm ch}_{1212}({\mathbf q},\omega =0)$, and (c) $-{\chi}^{\rm ch}_{1221}({\mathbf q},\omega =0)$ at $T=0.020$ eV, where orbitals 1 and 2 denote the $d_{x^2-y^2}$ and $d_{3z^2-r^2}$ orbitals, respectively.}
\label{figu2}
\end{figure*}

In Fig.~\ref{fig2}, we plot temperature dependence of $\lambda$, the maximum eigenvalue of the Eliashberg equation.
While $\lambda$ does not show appreciable temperature dependence in FLEX, 
$\lambda$ {is} drastically {enhanced} at low temperature $< 0.02$ eV  in TPSC. 
The characteristic enhancement of $\lambda$ in TPSC is attributed to  
the low-temperature behaviors of the spin and charge susceptibility.
In Fig. \ref{figu2}, we plot $\chi^{\rm sp}_{1111}({\mathbf q},\omega=0)$, $\chi^{\rm ch}_{1212}({\mathbf q},\omega=0)$, and $-\chi^{\rm ch}_{1221}({\mathbf q},\omega=0)$ at $T=0.020$ eV,
which indicate that the system has a strong incommensurate spin correlation in the $d_{x^2-y^2}$ orbital and strong {commensurate} inter-orbital fluctuations. 
{We here stress that there is no large peak in the charge susceptibilities
in the RPA and FLEX calculations, so that these enhanced inter-orbital charge fluctuations 
are purely due to the effects of vertex corrections. It should be noted that 
$\chi^{\rm ch}_{1221}(Q)$ has a negative peak around $Q = (\pi, \pi)$, which works 
as attractive force between the two orbitals for $d$-wave pairing just like antiferromagnetic spin fluctuations. [Note that the 
spin and charge sectors in Eq. (\ref{eliash-eq}) have opposite signs.] As we will see below, there is a close correlation between
the characteristic enhancement of $\lambda$ and the charge fluctuations.}
\begin{figure*}[htbp]
  \begin{center}
     \includegraphics[width=0.8\textwidth]{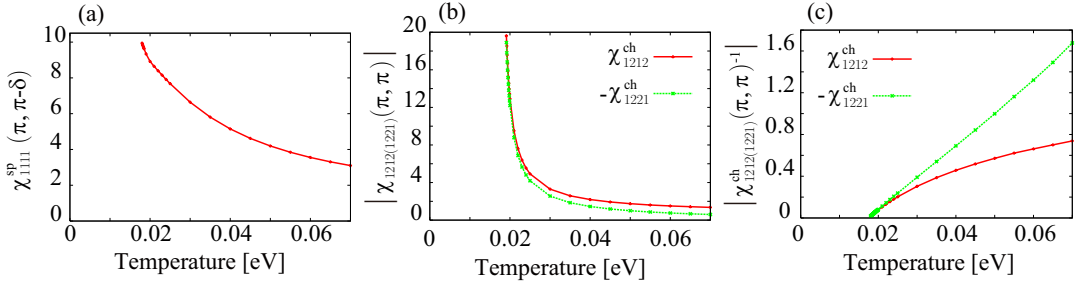}
  \end{center}
 \caption{(Color online) Temperature dependence of the {maximum value} in (a) ${\chi}^{\rm sp}_{1111}$, (b) $|{\chi}^{\rm ch}_{1212(1221)}|$, and (c) $|{{\chi}^{\rm ch}_{1212(1221)}}^{-1}|$.}
\label{lasusspin}
\end{figure*}

{Figure \ref{lasusspin} depicts temperature} dependence of the {maximum value} in the spin susceptibility, the orbital susceptibility, and {its inverse}. 
We can see that  the enhancement of the peaks in ${\chi}^{\rm ch}_{1212}$ and $-{\chi}^{\rm ch}_{1221}$ dominate over that of ${\chi}^{\rm sp}_{1111}$ for $T<0.02$ eV.
This behavior comes from the fact that  while ${\mathbf U}^{\rm sp}$ is renormalized substantially (${U}^{\rm sp}_{1111}\sim 1.16$ eV at $T=0.020$ eV), 
${\mathbf U}^{\rm ch}_{33}$ (${\mathbf U}^{\rm ch}_{44}$) becomes $\sim 1.55$ eV, which is {even} larger than the bare value 1.2 eV used in RPA and FLEX. 
{In fact, similar enhancement of charge channel is also} observed in the case of the single-orbital model.\cite{vilk1} 

\begin{figure}[htbp]
  \begin{center}
     \includegraphics[width=80mm]{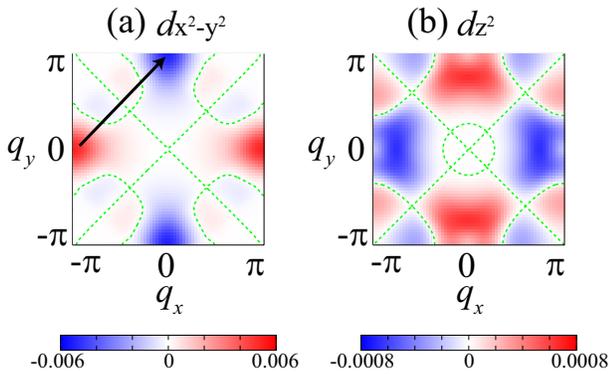}
  \end{center}
 \caption{(Color online) Gap function for the (a) $d_{x^2-y^2}$ orbital and the (b) $d_{3z^2-r^2}$ orbital at $T=0.022$ eV.  The gap functions have $d$-wave symmetry and the black arrow denotes the pair scattering mediated by antiferromagnetic spin fluctuations.}
\label{gap20}
\end{figure}

{Figure \ref{gap20} shows} the gap functions for the $d_{x^2-y^2}$ orbital and the $d_{3z^2-r^2}$ orbital ($\Delta_{11}$ and $\Delta_{22}$) 
at $T=0.022$ eV. {They} have the $d$-wave symmetry,
which is mediated by the {dominant} spin fluctuation, denoted by the black arrow. {This is the conventional situation where the orbital fluctuations remains small}.

\begin{figure}[htbp]
  \begin{center}
     \includegraphics[width=80mm]{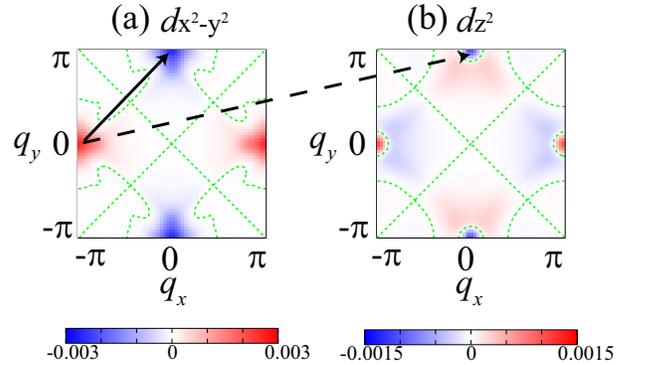}
  \end{center}
 \caption{(Color online) Plots similar to Fig.~\ref{gap20} for $T=0.018$ eV. The dashed black arrow denotes the pair scattering mediated by orbital fluctuations.}
\label{gap18}
\end{figure}

In Fig. \ref{gap18}, we plot the gap functions at {lower temperature} $T=0.018$ eV. 
{We see that characteristic structure emerges at $(\pi, 0)$ and $(0, \pi)$ in the 
$d_{3z^2-r^2}$ gap function, due to the inter-orbital effective interaction, dominantly
mediated by the orbital fluctuation $\chi^{\rm ch}_{1221}(q)$, which connects $\Delta_{11}$
and $ \Delta_{22}$ [Eq. (\ref{gapeq1})]. Since $\chi^{\rm ch}_{1221}(q)$ takes a large negative value 
around $(\pi,\pi)$, it cooperates with antiferromagnetic 
spin fluctuation to enhance the $d$-wave pairing instability.}

{This situation changes} in the stronger coupling regime ($U \sim 2.5$ eV), {where the system goes away from a superconducting instability}. 
\begin{figure*}[htbp]
  \begin{center}
     \includegraphics[width=0.8\textwidth]{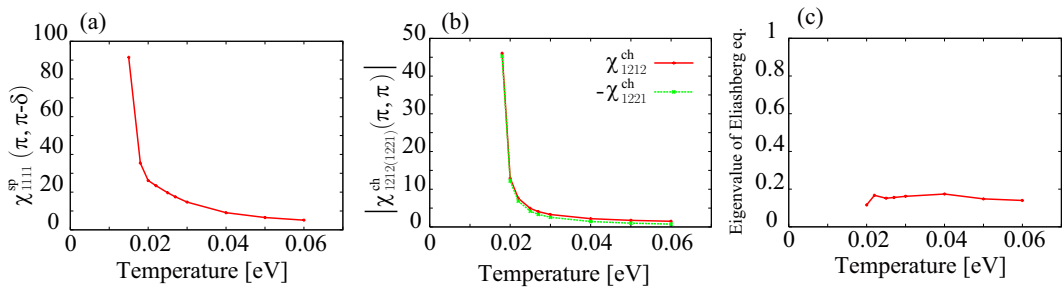}
  \end{center}
 \caption{(Color online) Temperature dependence of the {maximum value} in (a) ${\chi}^{\rm sp}_{1111}$, (b) $|{\chi}^{\rm ch}_{1212(1221)}|$, and (c) the maximum eigenvalue of the linearized Eliashberg equation obtained by TPSC for $U=2.5$ eV, $U'=2.0$ eV, and $J=0.25$ eV.}
\label{lasusu25}
\end{figure*}
This is because the dominant spin{/orbital} fluctuation{s} make the quasi-particle damping around $(0, \pi)$ and $(\pi, 0)$ significant, as is observed in the previous FLEX calculation.\cite{sakakibara1} Temperature dependence of the {maximum value} in the spin, orbital susceptibility and the maximum eigenvalue of the Eliashberg equation $\lambda$ for $U=2.5$ eV, $U'=2.0$ eV and $J=0.25$ eV are shown in Figs. \ref{lasusu25}(a), \ref{lasusu25}(b), and \ref{lasusu25}(c), respectively. 
{The} maximum value in ${\chi}^{\rm sp}_{1111}$ always dominates over that of ${\chi}^{\rm ch}_{1212}$ and $-{\chi}^{\rm ch}_{1212}$, and $\lambda$ does {not} show any enhancement.

\subsection{Iron-based superconductor: $\rm LaFeAsO$}
Let us now apply multi-orbital TPSC to the iron-based superconductor, LaFeAsO. 
The recent discovery of high $T_c$ superconductivity in F-doped LaFeAsO\cite{kamihara1} has stimulated a renewed interest in multi-orbital superconductors. 
As for the pairing mechanism of the iron-based superconductors, several scenarios have been proposed. Among them, the possibility of the sign-reversing $s_{\pm}$-wave superconductivity
mediated by spin fluctuations\cite{mazin1,kuroki1} have been extensively studied. While the $s_{\pm}$-wave solution {has been} obtained in the RPA\cite{kuroki1, ironRPA,graser01} or FLEX\cite{ironFLEX} calculation{s}
for the five-orbital $d$-model, recently, it has been proposed that vertex correction{s} {can} enhance orbital fluctuations, and   
the $s_{++}$-pairing without sign reversing becomes dominant.\cite{Kontani} 
In this subsection, we discuss how vertex correction{s} in TPSC affects superconductivity 
in the five-orbital $d$-model for LaFeAsO.

\begin{figure}[htbp]
  \begin{center}
     \includegraphics[width=65mm]{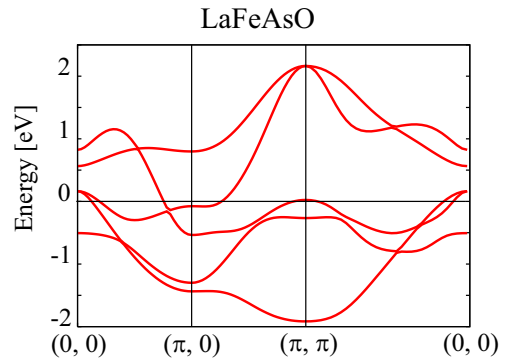}
  \end{center}
 \caption{(Color online) Band structure of the five-orbital $d$-model for LaFeAsO. The Fermi level is set at $0$ eV.}
   \label{fig:rotationgg}
\end{figure}

{The band} structure {of} the $d$-model is shown in Fig. \ref{fig:rotationgg}.
The bare coupling constants are set  to be {$U=1.5$ eV, $U'=1.2$ eV, and $J=0.15$ eV}.{\cite{comment2}} {The number of  electrons $n$ is 6.1}.
We employ 64 $\times$ 64 $k$-point meshes and 2048 Matsubara frequencies.
Hereafter orbitals 1, 2, 3, 4, and 5 denote the $d_{3z^2-r^2}$, $d_{xz}$, $d_{yz}$, $d_{x^2-y^2}$, and $d_{xy}$ orbitals, respectively.

\begin{figure*}[htbp]
  \begin{center}
     \includegraphics[width=0.8\textwidth]{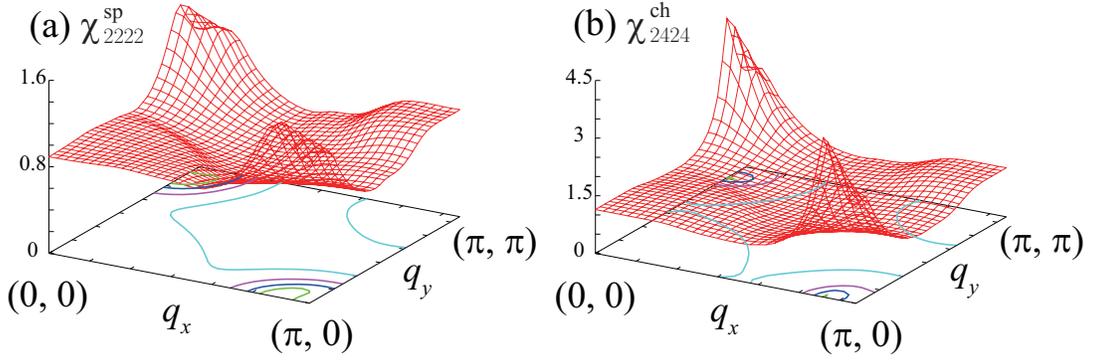}
  \end{center}
\caption{(Color online) (a) ${\chi}^{\rm sp}_{2222}({\mathbf q},\omega =0)$ and (b) ${\chi}^{\rm ch}_{2424}({\mathbf q},\omega =0)$  at $T=0.015$ eV, where orbitals 2 and 4 denote the $d_{xz}$ and $d_{x^2-y^2}$ orbitals, respectively.}
\label{fe-spi2}
\end{figure*}

\begin{figure}[htbp] 
  \begin{center}
     \includegraphics[width=65mm]{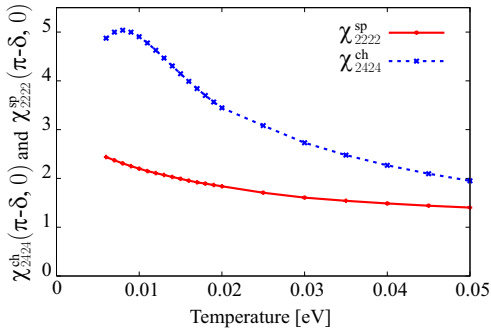}
  \end{center}
\caption{(Color online) Temperature dependence of the {maximum value} in ${\chi}^{\rm sp}_{2222}$ (red solid line) and ${\chi}^{\rm ch}_{2424}$ (blue dotted line) obtained by TPSC.}
\label{fe-el1}
\end{figure}

To see that TPSC {can} give {enhanced} orbital fluctuations, we plot $\chi^{\rm sp}_{2222}({\mathbf q},\omega=0)$ and $\chi^{\rm ch}_{2424}({\mathbf q},\omega=0)$ 
at $T=0.015$ eV in Figs. \ref{fe-spi2}(a) and \ref{fe-spi2}(b), respectively. 
{Clearly, both} susceptibilities have peaks around $(\pi, 0)$ and $(0, \pi)$, and
the peak in the orbital susceptibility is higher than that of the spin susceptibility.
Temperature dependence of these peaks are shown in Fig. \ref{fe-el1}.
We see that the peak of ${\chi}^{\rm ch}_{2424}({\mathbf q},\omega =0)$ {is} more drastically {enhanced} than that of ${\chi}^{\rm sp}_{2222}({\mathbf q},\omega =0)$ for $T>0.01$ eV.
While ${\chi}^{\rm ch}_{2424}({\mathbf q},\omega =0)$ has a broad maximum {peak} around $T=0.01$ eV,  ${\chi}^{\rm sp}_{2222}({\mathbf q},\omega =0)$ grows monotonously as temperature lowers. 

\begin{figure}[htbp] 
  \begin{center}
     \includegraphics[width=65mm]{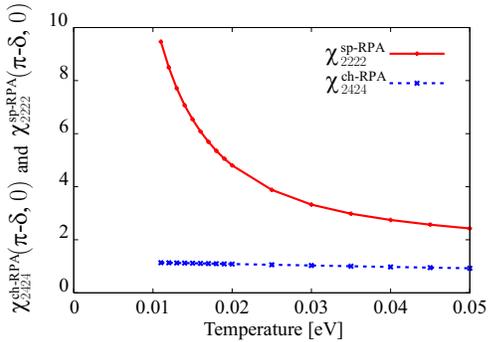}
  \end{center}
\caption{(Color online) Temperature dependence of the {maximum value} in ${\chi}^{\rm sp}_{2222}$ (red solid line) and ${\chi}^{\rm ch}_{2424}$ (blue dotted line) obtained by RPA. The bare coupling constants are set  to be $U=1.2$ eV, $U'=0.96$ eV, and $J=0.12$ eV.}
\label{rpas}
\end{figure}

The enhancement in the orbital susceptibility comes from the vertex correction in the charge susceptibility. 
To make this point clear, in Fig. \ref{rpas}, we plot temperature dependence of the {maximum value} in ${\chi}^{\rm ch}_{2424}({\mathbf q},\omega =0)$ and ${\chi}^{\rm sp}_{2222}({\mathbf q},\omega =0)$ obtained by RPA,
for which the bare coupling constants are set  to be $U=1.2$ eV, $U'=0.96$ eV, and $J=0.12$ eV. We see that while ${\chi}^{\rm sp}_{2222}({\mathbf q},\omega =0)$ diverges around $T=0.01$ eV,
${\chi}^{\rm ch}_{2424}({\mathbf q},\omega =0)$ has no significant temperature dependence.

\begin{figure}[htbp]
  \begin{center}
     \includegraphics[width=65mm]{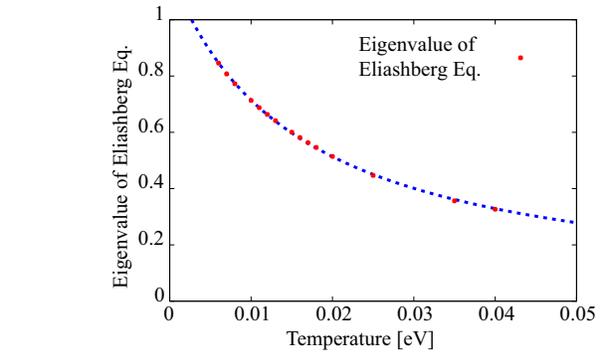}
  \end{center}
 \caption{(Color online) Temperature dependence of the maximum eigenvalue of the linearized Eliashberg equation. The blue dotted line is a guide to the eye.} 
\label{fe-el}
\end{figure}

In Fig. \ref{fe-el}, we show temperature dependence of the maximum eigenvalue of the Eliashberg equation. We see that the system has a superconducting transition around $T\sim 0.005$ eV.  

\begin{figure*}[htbp]
  \begin{center}
     \includegraphics[width=0.8\textwidth]{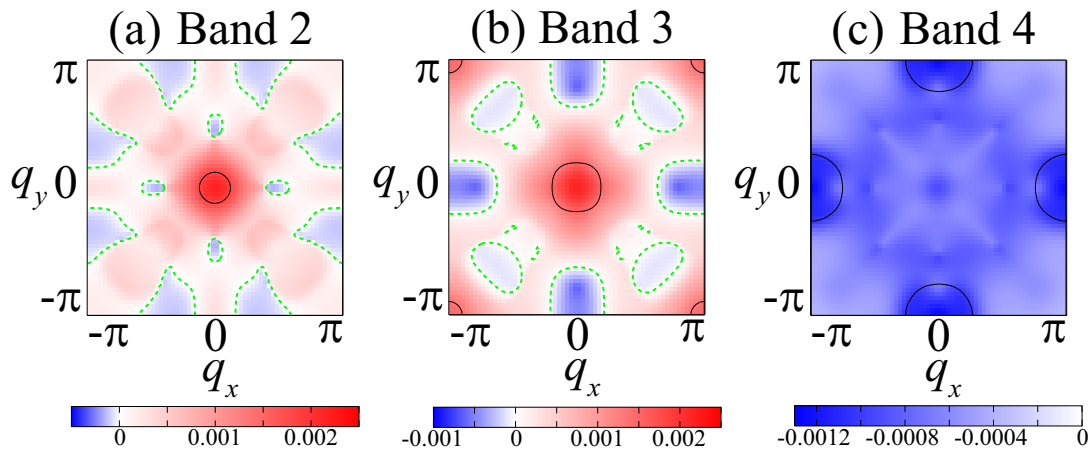}
  \end{center}
 \caption{(Color online) Gap functions obtained by TPSC for the bands with the (a) second, (b) third, and (c) fourth Kohn-Sham energy at $T=0.015$ eV. The black line and dotted green line represent the Fermi surface and nodes of gap functions, respectively.}  
\label{fegap3}
\end{figure*}

\begin{figure*}[htbp]
  \begin{center}
     \includegraphics[width=0.8\textwidth]{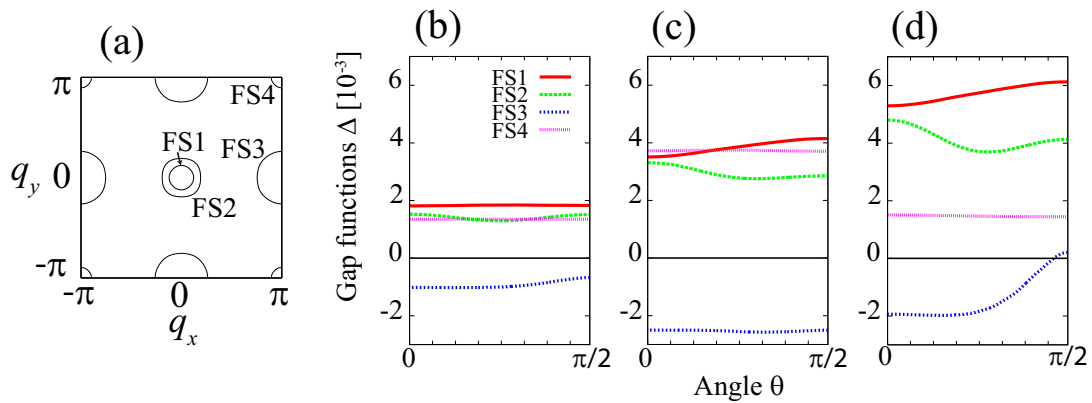}
  \end{center}
 \caption{(Color online) Gap function on the (a) Fermi surfaces by (b) TPSC, (c) RPA, and (d) TPSC without charge fluctuation, where $\theta$ is the rotation angle from the $k_y$ axis. }  
\label{rpavstpsc}
\end{figure*}

The associated eigenfunctions of the Eliashberg equation at $T=0.015$ eV (the gap functions) are shown in Fig.~\ref{fegap3} for the three bands crossing the Fermi level. 
We see that the{se} gap function{s have} the $s_\pm$ symmetry, indicating that the spin fluctuation is the {primary glue} of superconductivity. However, there is a {notable} difference between TPSC and RPA results in the amplitudes of the gap functions on the Fermi surface.
As we can see in Fig.~\ref{rpavstpsc}, the gap amplitude is large{r} for RPA than {TPSC}. This {indicates that} there is a frustration between the orbital-fluctuation-mediated pairing and 
the spin-fluctuation-mediated pairing.  Indeed, in TPSC, if we drop the contribution of the charge channel in the pairing interaction, namely consider only ${\mathbf V}(k)=-\frac{3}{2} {\mathbf U}^{\rm sp}{\boldsymbol \chi}^{\rm sp}(k)\ {\mathbf U}^{\rm sp}_0$ in the Eliashberg equation,  then we find that the gap amplitude becomes large on the Fermi surface as in the RPA {result} [see Fig.~\ref{rpavstpsc}(d)]. 

\section{Summary} \label{sec:summary}
To summarize, we have developed the two-particle self consistent method (TPSC) for the multi-orbital Hubbard model.  We derived self-consistent equations to determine vertex correction{s} in the spin and charge (orbital) susceptibiliti{es}. We applied this method to the effective models for $\rm La_2CuO_4$ and LaFeAsO.  We solved the linearized Eliashberg equation and found that vertex correction{s} play a crucial role in the multi-orbital superconductors.

In the two-orbital $d_{x^2-y^2}$-$d_{3z^2-r^2}$ model for $\rm La_2CuO_4$, while FLEX shows much lower $T_c$ than its experimental value $\sim 40$ K, {the} present TPSC {can} increase $T_c$ dramatically due to enhanced orbital fluctuation{s} via vertex corrections {for intermediate $U\sim$ 2.0} eV. 

In the iron-based superconductor LaFeAsO, we have studied whether orbital fluctuations {can be} enhanced  and induce the $s_{++}$-wave pairing within TPSC. {Indeed we have found that} some kinds of orbital fluctuations are enhanced {by considering vertex correction} and become even stronger than spin fluctuations.  However, their orbital fluctuations are not strong enough to {cause} the $s_{++}$-wave pairing, and the obtained gap function {has} the $s_{\pm}$ symmetry, although the gap magnitude is relatively suppressed due to a frustration between two kinds of pairing interactions mediated by spin and orbital fluctuations.
It is an interesting problem in future research whether the pairing symmetry changes for larger interaction parameters.
{Another important future issue is a systematic comparison between the present multi-orbital TPSC and other (diagrammatic) methods which consider the vertex corrections.}
Finally, we stress the importance of vertex corrections in multi-orbital systems for cooperative and competitive phenomena between spin and orbital degree{s} of freedoms.

\begin{acknowledgements}
We thank H. Kontani and S. Onari for stimulating discussions. This work was supported by Grants-in-Aid for Scientific Research (No.~23340095) from MEXT and JST-PRESTO, Japan.
\end{acknowledgements}

\appendix
\section{Definition of correlation functions} \label{sec:appendix1}
In the single-orbital case, correlation functions for spin $S^z({\mathbf r})$ and charge $n({\mathbf r})$ are defined as
\begin{subequations}
\begin{align}
{\chi}^{\rm sp}(1,2)&=\langle  T_{\tau} S^z(1)S^z(2) \rangle , \\
{\chi}^{\rm ch}(1,2)&=\langle  T_{\tau} n(1)n(2) \rangle - \langle n(1) \rangle \langle n(2) \rangle ,
\end{align}
\end{subequations}
where an abbreviation $1=({\mathbf r}_1, {\tau}_1)$ denotes a position ${\bf r}_1$ and an imaginary time $\tau_1$, and $T_{\tau}$ is the time ordering operator.  Time dependence of a generic operator $Q({\mathbf r})$ is defined as $Q({\mathbf r}, \tau )=e^{-\tau H} Q({\mathbf r}) e^{\tau H}$.  ${\chi}^{\rm sp (ch)}(q)$ in the main text is the Fourier transform of the above real-space representation.  The two sum rules (Eqs.~(\ref{eq:sumrule})), which play a central role in TPSC, originate from the definition at equal time, that is, ${\chi}^{\rm sp (ch)}(1,1^+)$ with $1^+=({\mathbf r}_1,{\tau}_1+\delta)$ ($\delta > 0$). 

In the multi-orbital case, we consider correlation functions for ${\mathbf S}_{\mu \nu}(1)=(c_{\mu \uparrow}^\dagger(1), c_{\nu \downarrow}^\dagger(1))\,{\boldsymbol \sigma}\,(c_{\mu \uparrow}(1), c_{\nu \downarrow}(1))^T$, and $n_{\mu \nu}(1)=c_{\mu \uparrow}^{\dagger}(1)c_{\nu \uparrow}(1)+c_{\mu \downarrow}^{\dagger}(1)c_{\nu \downarrow}(1)$.  These correlation functions for spin and charge channels are defined as
\begin{subequations}
\begin{align}
{\chi}^{{\rm sp}z}_{\lambda \mu \nu \xi}(1,2)&=\langle  T_{\tau} {S^z}_{\lambda \mu}(1) {S^z}_{\xi\nu}(2) \rangle, \\
{\chi}^{{\rm sp}\pm}_{\lambda \mu \nu \xi}(1,2)&=\langle  T_{\tau} {S^+}_{\lambda \mu}(1) {S^-}_{\xi\nu}(2) \rangle, \\
{\chi}^{\rm ch}_{\lambda \mu \nu \xi}(1,2)&=\langle  T_{\tau} n_{\lambda \mu}(1) n_{\xi\nu}(2) \rangle - \langle n_{\lambda \mu }(1) \rangle \langle n_{\xi\nu}(2) \rangle.
\end{align}
\end{subequations}

The sum rules of Eqs.~(\ref{eq:rule1}) and (\ref{eq:rule2}) come from the following definitions at equal time,
\begin{widetext}
\begin{subequations}
\begin{align}
\begin{split}
{\chi}^{{\rm sp}z}_{\lambda \mu \nu \xi}(1,1^+)&=\langle (c^{\dagger}_{\lambda \uparrow}(1)c_{\mu \uparrow}(1)-c^{\dagger}_{\lambda \downarrow}(1)c_{\mu \downarrow}(1))(c^{\dagger}_{\xi \uparrow}(1^+)c_{\nu \uparrow}(1^+)-c^{\dagger}_{\xi \downarrow}(1^+)c_{\nu \downarrow}(1^+))\rangle \\
&-\langle (c^{\dagger}_{\lambda \uparrow}(1)c_{\mu \uparrow}(1)-c^{\dagger}_{\lambda \downarrow}(1)c_{\mu \downarrow}(1)) \rangle \langle (c^{\dagger}_{\xi \uparrow}(1^+)c_{\nu \uparrow}(1^+)-c^{\dagger}_{\xi \downarrow}(1^+)c_{\nu \downarrow}(1^+))\rangle
\end{split}
\\
{\chi}^{{\rm sp}\pm}_{\lambda \mu \nu \xi}(1,1^+)&=2\langle c^{\dagger}_{\lambda \uparrow}(1)c_{\mu \downarrow}(1)c^{\dagger}_{\xi \downarrow}(1^+)c_{\nu \uparrow}(1^+)\rangle 
\\
\begin{split}
{\chi}^{\rm ch}_{\lambda \mu \nu \xi}(1,1^+)&=\langle (c^{\dagger}_{\lambda \uparrow}(1)c_{\mu \uparrow}(1)+c^{\dagger}_{\lambda \downarrow}(1)c_{\mu \downarrow})(c^{\dagger}_{\xi \uparrow}(1^+)c_{\nu \uparrow}(1^+)+c^{\dagger}_{\xi \downarrow}(1^+)c_{\nu \downarrow}(1^+))\rangle \\
&-\langle (c^{\dagger}_{\lambda \uparrow}(1)c_{\mu \uparrow}(1)+c^{\dagger}_{\lambda \downarrow}(1)c_{\mu \downarrow})(c^{\dagger}_{\xi \uparrow}(1^+)c_{\nu \uparrow}(1^+)+c^{\dagger}_{\xi \downarrow}(1^+)c_{\nu \downarrow}(1^+))\rangle. 
\end{split}
\end{align}
\end{subequations}
\end{widetext}

\section{Ansatz for effective interactions in single-orbital case} \label{sec:appendix2}
Following Vilk and Tremblay,\cite{vilk1} let us derive the ansatz, Eq.~(\ref{eq:ansatz}), used in TPSC calculations.
The four-point vertex function, ${\Gamma}_{\sigma{\sigma}'} $, between electrons with spin $\sigma$ and ${\sigma}'$ is given by
\begin{equation}
{\Gamma}_{\sigma{\sigma}'}\delta (1-3)\delta(2-4)\delta(2-1^+)=\frac{\delta {\Sigma}_{\sigma}(1,2)}{\delta G_{{\sigma}'}(3,4)} , \label{eq:vertex}
\end{equation}
where $G_{\sigma}(1,2)$ and ${\Sigma}_{\sigma}(1,2)$ is the dressed Green's function and self energy with spin $\sigma$. 
The equation of motion and the Dyson equation leads to the relation,
\begin{equation}
\begin{split}
& {\Sigma}_{\sigma}(1,\bar{1})G_{\sigma}(\bar{1},2) \\
& ~~~~= -U\langle T_{\tau}[ {c}_{\bar{\sigma}}^{\dagger}(1^{++}){c}_{\bar{\sigma}}(1^{+}){c}_{\sigma}(1){c}_{\sigma}^{\dagger}({2})]\rangle,
\end{split}
\end{equation}
with $\bar{\sigma}=-\sigma$.  Here a bar over a number means the integral over position and imaginary time.
The four-point correlation function can be approximated by the local correlation function and the Green's function as follows,
\begin{equation}
\begin{split}
& -U\langle T_{\tau}[ {c}_{\bar{\sigma}}^{\dagger}(1^{++}){c}_{\bar{\sigma}}(1^{+}){c}_{\sigma}(1){c}_{\sigma}^{\dagger}({2})]\rangle  \\
& ~~~~ \sim U\frac{\langle n_\uparrow (1)n_\downarrow (1)\rangle}{\langle n_\uparrow (1) \rangle \langle n_\downarrow (1) \rangle}G_{\bar{\sigma}}(1,1^+)G_{\sigma}(1,{2}). \label{eq:app}
\end{split}
\end{equation}
By substituting Eq.~(\ref{eq:app}) into Eq.~(\ref{eq:vertex}), we can obtain 
\begin{widetext}
\begin{equation}
\begin{split}
{\Gamma}_{\sigma{\sigma}'}\delta (1-3)\delta(2-4)\delta(2-1^+)&=\frac{\delta {\Sigma}_{\sigma}(1,2)}{\delta G_{{\sigma}'}(3,4)} 
=\frac{ \delta \left[ U\frac{\langle n_{\uparrow}n_{\downarrow} \rangle}{\langle n_{\uparrow} \rangle \langle n_{\downarrow}\rangle} G_{\bar{\sigma}}(1,1^+)\delta(1-2) \right] }{\delta G_{\sigma '}(3,4)} \\
&=\frac{ \delta \left[ U\frac{\langle n_{\uparrow}n_{\downarrow} \rangle}{\langle n_{\uparrow} \rangle \langle n_{\downarrow}\rangle}  \right] }{\delta G_{\sigma '}(3,4)} G_{\bar{\sigma}}(1,1^+)\delta(1-2)+U\frac{\langle n_{\uparrow}n_{\downarrow} \rangle}{\langle n_{\uparrow} \rangle \langle n_{\downarrow}\rangle}\frac{ \delta G_{\bar{\sigma}}(1,1^+) }{\delta G_{\sigma '}(3,4)}\delta(1-2).
\end{split}
\end{equation}
\end{widetext}
The last term of this equation is proportional to ${\delta}_{\bar{\sigma} \sigma '}$ via 
\begin{equation}
\frac{\delta G_{\bar{\sigma}}(1,1^+)}{\delta G_{\sigma '}(3,4)}=\delta_{\bar{\sigma} \sigma '}\delta(1-3)\delta(4-1^+),
\end{equation}
and then contributes to the spin channel, $U_{\rm sp}=\Gamma_{\sigma \bar{\sigma}}-\Gamma_{\sigma \sigma}$.
This just provides Eq.~(\ref{eq:ansatz}) for $U_{\rm sp}$.

\section{Ansatz in multi-orbital systems} \label{sec:appendix3}
In this section, let us extend the above-mentioned ansatz into the multi-orbital case.
In this case, the four-point vertex function has orbital indices, $\lambda$, $\mu$, $\nu$, $\xi$ besides spin index, $\sigma$.
We here consider $\Gamma_{(\mu\mu\sigma)(\nu\nu\sigma')}$, which can be written by only orbital-diagonal components.
\begin{widetext}
\begin{align}
&{\Gamma}_{(\mu \mu\sigma) (\nu \nu{\sigma}')}\delta(1-3)\delta(2-4)\delta(2-1^+)=\frac{\delta {\Sigma}_{\mu \mu\sigma}(1,2)}{\delta G_{\nu \nu{\sigma}'}(3,4)}
= \frac{\delta \left[ {\Sigma}_{\mu \mu\sigma}(1,\bar{5}) [{\mathbf G} (\bar{5},\bar{6}) {\mathbf G^{-1}}(\bar{6},2) ]_{\mu \mu\sigma} \right] }{\delta G_{\nu \nu{\sigma}'}(3, 4)} \nonumber \\
&\sim\frac{\delta}{\delta G_{\nu \nu\sigma '}(3, 4)} 
\left( 
-U\frac{\langle n_{\mu {\sigma}}n_{\mu \bar{\sigma}}\rangle}{\langle n_{\mu {\sigma}} \rangle \langle n_{\mu \bar{\sigma}} \rangle }G_{\mu \mu\bar{\sigma}}(1, 1^+)
-\sum_{\xi \neq \mu}U'\frac{\langle n_{\mu \sigma}n_{\xi \bar{\sigma}}\rangle}{\langle n_{\mu \sigma} \rangle \langle n_{\xi \bar{\sigma}} \rangle }G_{\xi \xi \bar{\sigma}}(1, 1^+) 
-\sum_{\xi \neq \mu}(U'-J)\frac{\langle n_{\mu \sigma}n_{\xi \sigma}\rangle}{\langle n_{\mu \sigma} \rangle \langle n_{\xi \sigma} \rangle }G_{\xi \xi \sigma}(1, 1^+)
 \right) \nonumber \\
&\sim
-U\frac{\langle n_{\mu {\sigma}}n_{\mu \bar{\sigma}}\rangle}{\langle n_{\mu {\sigma}} \rangle \langle n_{\mu \bar{\sigma}} \rangle }\frac{\delta G_{\mu \mu\bar{\sigma}}(1, 1^+)}{\delta G_{\nu \nu \sigma '}(3, 4)}
-\sum_{\xi \neq \mu}U'\frac{\langle n_{\mu \sigma}n_{\xi \bar{\sigma}}\rangle}{\langle n_{\mu \sigma} \rangle \langle n_{\xi \bar{\sigma}} \rangle }\frac{\delta G_{\xi \xi \bar{\sigma}}(1, 1^+)}{\delta G_{\nu \nu \sigma '}(3, 4)}
-\sum_{\xi \neq \mu}(U'-J)\frac{\langle n_{\mu \sigma}n_{\xi \sigma}\rangle}{\langle n_{\mu \sigma} \rangle \langle n_{\xi \sigma} \rangle }\frac{\delta G_{\xi \xi \sigma}(1, 1^+)}{\delta G_{\nu \nu \sigma '}(3, 4)}.
\label{eq:ansatz2}
\end{align}
\end{widetext}
Here, following the single-orbital case, we have introduced the following approximations, 
\begin{align*}
{\Sigma}_{\mu \mu \sigma}G_{\mu \mu \sigma}G_{\mu \mu \sigma}^{-1}&\sim -U\frac{\langle n_{\mu {\sigma}}n_{\mu \bar{\sigma}}\rangle}{\langle n_{\mu {\sigma}} \rangle \langle n_{\mu \bar{\sigma}} \rangle }G_{\mu \mu \bar{\sigma}}(1, 1^+), \\
{\Sigma}_{\mu \mu \sigma}G_{\xi \xi \sigma}G_{\xi \xi \sigma}^{-1}&\sim -U'\frac{\langle n_{\mu {\sigma}}n_{\xi \bar{\sigma}}\rangle}{\langle n_{\mu {\sigma}} \rangle \langle n_{\xi \bar{\sigma}} \rangle }G_{\xi \xi\bar{\sigma}}(1, 1^+), \\
{\Sigma}_{\mu \mu \sigma}G_{\xi \xi \bar{\sigma}}G_{\xi \xi \bar{\sigma}}^{-1}&\sim -U'\frac{\langle n_{\mu {\sigma}}n_{\xi \bar{\sigma}}\rangle}{\langle n_{\mu {\sigma}} \rangle \langle n_{\xi \bar{\sigma}} \rangle }G_{\xi \xi {\sigma}}(1, 1^+).
\end{align*}
Eq.~(\ref{eq:re1}) for $U^{\rm sp}_{\mu \mu \mu \mu}$ can be obtained from the first term of Eq.~(\ref{eq:ansatz2}), since the intra-orbital Coulomb interaction for the orbital $\mu$ is proportional to ${\delta}_{\mu \mu}{\delta}_{\sigma \bar{\sigma}}$. 
In the same way, Eq.~(\ref{eq:re2}) for $U^{\rm sp}_{\mu \nu \mu \nu}$ and Eq.~(\ref{eq:re3}) for $U^{\rm sp}_{\mu \nu \mu \nu}-J^{\rm sp}$ with $\mu \ne \nu$ can be obtained from the second and the third terms in Eq.~(\ref{eq:ansatz2}), respectively.


\begin{thebibliography}{99}
\bibitem{comment} The Hund's coupling in the charge channel is always renormalized to be small in our case, and sometimes causes unexpected problematic behavior.
\bibitem{comment2}{Even if we change the values of these interaction parameters by 10 \%, the qualitative tendency of the present result does not change.}
\end{thebibliography}
\end{document}